\documentclass[english,pra,twocolumn]{revtex4-1}
\usepackage[T1]{fontenc}
\usepackage[latin9]{inputenc}
\usepackage{color}
\usepackage{amssymb}
\usepackage{graphicx}

\makeatletter

\@ifundefined{textcolor}{}
{%
 \definecolor{BLACK}{gray}{0}
 \definecolor{WHITE}{gray}{1}
 \definecolor{RED}{rgb}{1,0,0}
 \definecolor{GREEN}{rgb}{0,1,0}
 \definecolor{BLUE}{rgb}{0,0,1}
 \definecolor{CYAN}{cmyk}{1,0,0,0}
 \definecolor{MAGENTA}{cmyk}{0,1,0,0}
 \definecolor{YELLOW}{cmyk}{0,0,1,0}
}

\makeatother

\usepackage{babel}
\begin{document}

\title{Josephson oscillations and self-trapping of superfluid fermions \\ in a double-well potential}

\author{Peng Zou}
\affiliation{INO-CNR BEC Center and Dipartimento di Fisica, Universit\`a di Trento, 38123 Povo, Italy}
\author{Franco Dalfovo}
\affiliation{INO-CNR BEC Center and Dipartimento di Fisica, Universit\`a di Trento, 38123 Povo, Italy}

\begin{abstract}
We investigate the behaviour of a two-component Fermi superfluid in a double-well
potential. We numerically solve the time dependent Bogoliubov-de Gennes equations and
characterize the regimes of Josephson oscillations and self-trapping for different potential
barriers and initial conditions. In the weak link limit the results agree with a two-mode model
where the relative population and the phase difference between the two wells obey coupled
nonlinear Josephson equations. A more complex dynamics is predicted for large amplitude
oscillations and large tunneling.
\end{abstract}


\date{January 9, 2014}

\maketitle

\section{Introduction}

The Josephson effect \cite{josephson,barone} is one of the key features
of superconductors and superfluids. It involves very fundamental properties
of these systems and has important applications. The physics of the Josephson
junctions can be effectively investigated with ultracold gases confined in a double-well
potential \cite{leggett,book,gati}. The case of weakly linked Bose-Einstein condensates
(BEC) has been widely studied. In the seminal papers of  Smerzi {\it et al.} \cite{smerzi},
coupled nonlinear Josephson equations for the relative population and the
phase difference between the two wells were derived by assuming the system to be
described by a superposition of left and right localized condensates (two-mode model)
governed by the Gross-Pitaevskii (GP) equation (see also \cite{two-mode}). Such
nonlinear Josephson equations
admit solutions in the form of small periodic oscillations, whose period is
determined by two key parameters: the mean-field (on-site) energy and the
tunnelling energy. When the nonlinearity arising from the mean-field interaction
exceeds a critical value, the system may exhibit self-trapped solutions with
the relative population oscillating around a nonzero value.  A large number of
theoretical papers have been published along this line and experiments have also
been performed \cite{gati,florence,oberthaler,steinhauer,leblanc}.

Much less is known about Josephson effects in dilute Fermi gases. The
Bogoliubov-de Gennes (BdG) equations for a two-component superfluid in the
crossover from the Bardeen-Cooper-Schrieffer (BCS) phase to BEC  were used
in Ref.~\cite{strinati} to describe a stationary supercurrent flowing in the presence
of a three-dimensional barrier with a slab geometry; the current-phase relation
and the critical current were studied in the crossover for relatively low barriers,
i.e., height of the barrier smaller than the chemical potential of the superfluid. The
same problem was also investigated by means of a density functional approach
describing bosonic Cooper pairs \cite{salasnich}; the equation of state of the gas
was included via a suitable parametrization and the order parameter of the superfluid
was obtained as the solution of a nonlinear Schr\"odinger equation (NLSE). This method
gives results in good agreement with the BdG results of Ref.~\cite{strinati} from unitarity
to the BEC limit.  For a double-well potential in the weak link limit (i.e., large
barriers) the same density functional can be used to derive coupled nonlinear
equations for the relative population and the phase difference analog to
those for BECs \cite{salasnich2}. A similar NLSE has been used to discuss in
detail the transition from Josephson oscillations to self-trapping \cite{adhikari}.

Some open issues are worth considering. First, the applicability
of a two-mode model to weakly linked dilute Fermi superfluids has been tested
so far only within a density functional approach describing a gas of bosonic
pairs (namely Cooper pairs, which become molecules in the BEC limit);
being a generalization of the GP equation, the theory naturally reduces to
the two-mode model under the same assumptions as for coupled BECs. It is thus
interesting to test the two-mode model also within a more microscopic theory
like BdG which includes fermionic degrees of freedom. Second, the
available BdG calculations \cite{strinati} and their comparison with the density
functional results \cite{salasnich} are limited to the case of a stationary
current through a low and thick barrier, where the flow is almost hydrodynamic
and a local density approximation can be applied \cite{watanabe}; time
dependent simulations with higher and thinner barriers can provide a
a more stringent and informative test. Finally, the stationary BdG equations
does not include bosonic collective modes (e.g., phonons) in the spectrum of
excitations and cannot address the problem of dynamical instabilities, soliton
nucleation, phase slips, etc., which may occur in a superfluid flow in the presence
of a potential barrier. This type of physics can instead be addressed by
time-dependent BdG simulations.

Motivated by the above arguments, we numerically solve the time-dependent
BdG equations of a superfluid gas of fermions confined in a box with a square
potential barrier at the center. The square barrier is a convenient choice for
computational reasons, but the main results of this work would not change
by using barriers of different shape. In the limit of weakly linked superfluids we find
periodic oscillations whose frequency approaches the prediction of the two-mode
model, namely the Josephson ``plasma''  frequency $\hbar \omega_p=\sqrt{E_C E_J}$
\cite{book}, where the tunneling energy $E_J$ and the on-site mean-field energy $E_C$
are obtained by solving the stationary BdG equations in the same configurations.
By increasing the population imbalance we explore the transition from Josephson
oscillations to self-trapping. We compare the results with those obtained by solving
the NLSE of Refs.~\cite{salasnich,salasnich2,adhikari}, as well as with the predictions of the coupled
nonlinear Josephson equations which can be derived from the NLSE in the
weak link limit. The latter equations turn out to agree surprisingly well with the BdG
simulations, provided the parameters $E_J$ and $E_C$ are consistently calculated
within the same BdG theory.  Finally, for large amplitude oscillations and lower barriers
the dynamics is complex, involving a combination of Josephson-like oscillations,
phonons and solitons.

\section{Bogoliubov-de Gennes equations and system configuration}

We consider a three-dimensional atomic Fermi gas at zero temperature with
equal populations of two spin components. The interaction between atoms in
different spin states is characterized by the s-wave scattering length $a$. In
experiments, this parameter can be tuned from small negative (BCS regime)
to small positive (BEC regime) values by applying an external magnetic field
through a Feshbach resonance. On resonance the scattering length diverges
and the Fermi gas manifests universal properties (unitarity) \cite{rmp08}. We
describe the gas in the BCS-BEC crossover by means of the BdG equations \cite{BdG}
\begin{equation}
\left[\begin{array}{cc}
\hat{H} & \Delta(r)\\
\Delta^{*}({\bf r}) & -\hat{H}
\end{array}\right]\left[\begin{array}{c}
u_{\eta}({\bf r})\\
v_{\eta}({\bf r})
\end{array}\right]=\varepsilon_{\eta}\left[\begin{array}{c}
u_{\eta}({\bf r})\\
v_{\eta}({\bf r})
\end{array}\right]
\label{statBdG}
\end{equation}
where the functions $u_{\eta}$ and $v_{\eta}$ are the fermionic quasiparticle
amplitudes and $\varepsilon_{\eta}$ the corresponding quasiparticle eigenenergy;
$\hat{H}=-\hbar^{2}\nabla^{2}/(2m)+V_{\rm ext}({\bf r})-\mu$ is a single-particle
grand-canonical Hamiltonian for atoms of mass $m$ subject to an external potential
$V_{\rm ext}$, and $\mu$ is the chemical potential. The quasiparticle amplitudes
obey the normalization condition $\int d^3r \left[u_\eta^*({\bf r})u_{\eta'}({\bf
r}) +v_\eta^*({\bf r})v_{\eta'}({\bf r})\right]=\delta_{\eta,\eta'}$. The order parameter
of the superfluid phase is $\Delta({\bf r})=-\sum_{\eta} g_{\rm eff}({\bf r})u_{\eta}({\bf r})
v_{\eta}^{*}({\bf r})$, where $g_{\rm eff}$ is a coupling constant which accounts for the
interaction between atoms (see Eq.~(\ref{geff})); finally, the atom density is $n({\bf r})=2\sum_{\eta}
\left|v_{\eta}({\bf r})\right|^{2}$. We also solve the time-dependent version of the
BdG equations \cite{challis}:
\begin{equation}
\left[\begin{array}{cc}
\hat{H} & \Delta({\bf r},t)\\
\Delta^{*}({\bf r},t) & -\hat{H}
\end{array}\right]\left[\begin{array}{c}
u_{\eta}({\bf r},t)\\
v_{\eta}({\bf r},t)
\end{array}\right]=i\hbar\frac{\partial}{\partial t}\left[\begin{array}{c}
u_{\eta}({\bf r},t)\\
v_{\eta}({\bf r},t)
\end{array}\right] \; .
\label{tdbdg}
\end{equation}

We use an external potential $V_{\rm ext}(x)$ which depends on the longitudinal
coordinate $x$ only, consisting of two hard walls placed at $x=\pm L/2$ and a
rectangular barrier of height $V_0$ and width $d$ centered at $x=0$. The lengths
$L$ and $d$ are such that the density and the order parameter of the superfluid
are almost constant in the central region of each potential well on the left and
right sides of the barrier.  In the transverse directions the system is assumed to be
uniform; in practice, we solve the equations in a transverse square box of size
$L_\perp=13k_F^{-1}$  imposing periodic boundary conditions for all functions
$u$ and $v$; we checked that this value of $L_\perp$ is large enough to make
finite size effects negligible.  The average density $n_0$ of the ground state
in this potential can be used to define the Fermi wave vector of noninteracting
fermions of that density, $k_{F}=(3\pi^{2}n_0)^{1/3}$. We use $k_F^{-1}$ as
the unit of length; we also use the corresponding Fermi energy,
$E_{F}=\hbar^{2}k_{F}^{2}/(2m)$, as energy unit and $\hbar/E_F$ as time
unit. In the calculations we can either fix the total
number of atoms in the box or the density at a point at will.

The BdG equations are solved by using the same numerical method as in
\cite{scott}. In particular, the solution of the stationary equations (\ref{statBdG})
are found by a self-consistent iterative procedure. From the solution we can
calculate the grand canonical energy $E=\langle \hat{H} - \mu \hat{N} \rangle$
of the system as
\begin{equation}
E = \int \! d{\bf r} \sum_\eta [ 2(\mu-\varepsilon_\eta )|v_\eta({\bf r})|^2
+ \Delta^*({\bf r}) u_\eta({\bf r})v_\eta^*({\bf r})] \; .
\label{eq:energydens}
\end{equation}
The time-dependent equations (\ref{tdbdg}) are instead integrated by means
of a $4$-th order Runge-Kutta algorithm.

The BdG equations require a regularization procedure to cure ultraviolet
divergences. Here we use the method suggested in Ref.~\cite{bulgac}. For this,
we choose a cutoff energy $E_{\rm cut}$ sufficiently far above the Fermi
energy. Then, for a given external potential $V_{\rm ext}({\bf r})$ and
chemical potential $\mu$, we define a local Fermi wave vector $k_F({\bf r})$
from the relation $\mu=\hbar^{2}k_{F}^2({\bf r})/(2m)+V_{\rm ext}({\bf r})$
and a cutoff wave vector $k_{\rm cut}$ from $E_{\rm cut}  =
\hbar^{2}k_{\rm cut}^2 ({\bf r})/(2m)+V_{\rm ext}({\bf r})-\mu$. Finally, the
regularization of the interaction consists of replacing the bare coupling constant
$g=4\pi\hbar^2 a/m$, in the definition of $\Delta$ with an effective $g_{\rm eff}$
given by
\begin{equation}
\frac{1}{g_{\rm eff}({\bf r})}=\frac{1}{g}-\frac{mk_{\rm cut}({\bf
r})}{2\pi^{2}\hbar^{2}}
\left[1-\frac{k_{F}({\bf r})}{2k_{\rm cut}({\bf r})}\ln \frac{k_{\rm cut}
({\bf r})+k_{F}({\bf r})}{k_{\rm cut}({\bf r}) - k_{F}({\bf r})}\right] \;  .
\label{geff}
\end{equation}
The cutoff energy, $E_{\rm cut} $ is chosen large enough
to ensure the convergence to cutoff independent results.

\section{Dynamics of weakly linked Fermi superfluids at unitarity}

\begin{figure}
\includegraphics[scale=0.3]{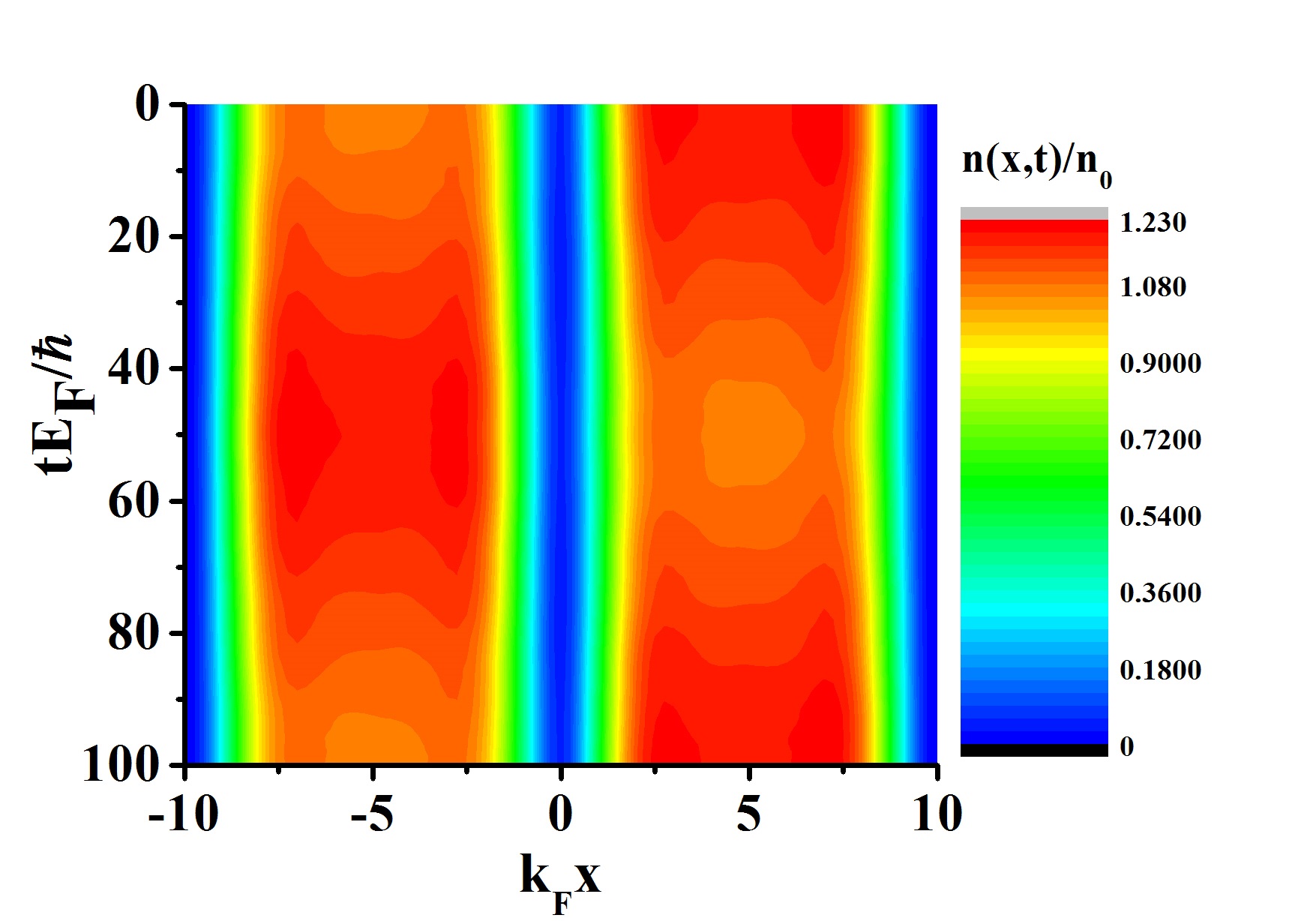}
\caption{Evolution of the density distribution $n(x,t)/n_0$ of a two-component
superfluid Fermi gas at unitarity and zero temperature obtained by solving
the time-dependent BdG equations (\ref{tdbdg}).  Time, in units of $\hbar/E_F$,
flows from top to bottom.  The gas is uniform
in the transverse directions and confined between hard walls in the longitudinal
direction at $x=\pm L/2$ with $L=20 k_F^{-1}$, with a central square
barrier of height $V_0=5 E_F$ and width $d=0.6 k_F^{-1}$ respectively.
The number of atoms is $N=100$. The initial imbalance is
produced by adding a constant offset potential $V_{\rm off}$ at $t<0$ on
the left side only; here we use $V_{\rm off}=0.05E_F$, which corresponds
to an initial relative imbalance $z_0= (N_L-N_R)/N=-0.06$. At $t=0$ the
offset potential is removed and the system is let to evolve in time.  }
\label{figjosephson0}
\end{figure}

Let us consider fermions at unitarity ($1/(k_Fa)=0$) in the presence
of a thin ($d \sim k_F^{-1}$) and high ($V_0>\mu$) square barrier centered at
$x=0$.  We can define the number of atoms on the left, $N_L$, and right, $N_R$,
as the integrals of the atom density $n(x)$ separately in the two regions of negative
and positive $x$, respectively. The relative population imbalance can be defined
as  $z = (N_L-N_R)/N$, where $N=N_L+N_R$. Another key quantity is
the phase $\phi(x)$  of the complex order parameter $\Delta(x)$, which can
also be different in the two wells. We define the right and left phases as
$\phi_{R}=\phi(x= L/4)$ and  $\phi_{L}=\phi(x=-L/4)$ respectively, and the phase
difference as $\Phi = \phi_R -\phi_L$.

Our simulations start from an imbalanced configuration with $z_0 \equiv z(t=0) \neq 0$.
This is obtained by first solving the stationary BdG equations (\ref{statBdG}) with a small
constant offset potential $V_{\rm off}$ on the left side of the barrier. The ground state
solution in such an asymmetric potential is then used as the initial ($t=0$)
state in the integration of the time-dependent BdG equations (\ref{tdbdg}) in the
symmetric double-well, after removing  $V_{\rm off}$.

\begin{figure}
\includegraphics[scale=0.3]{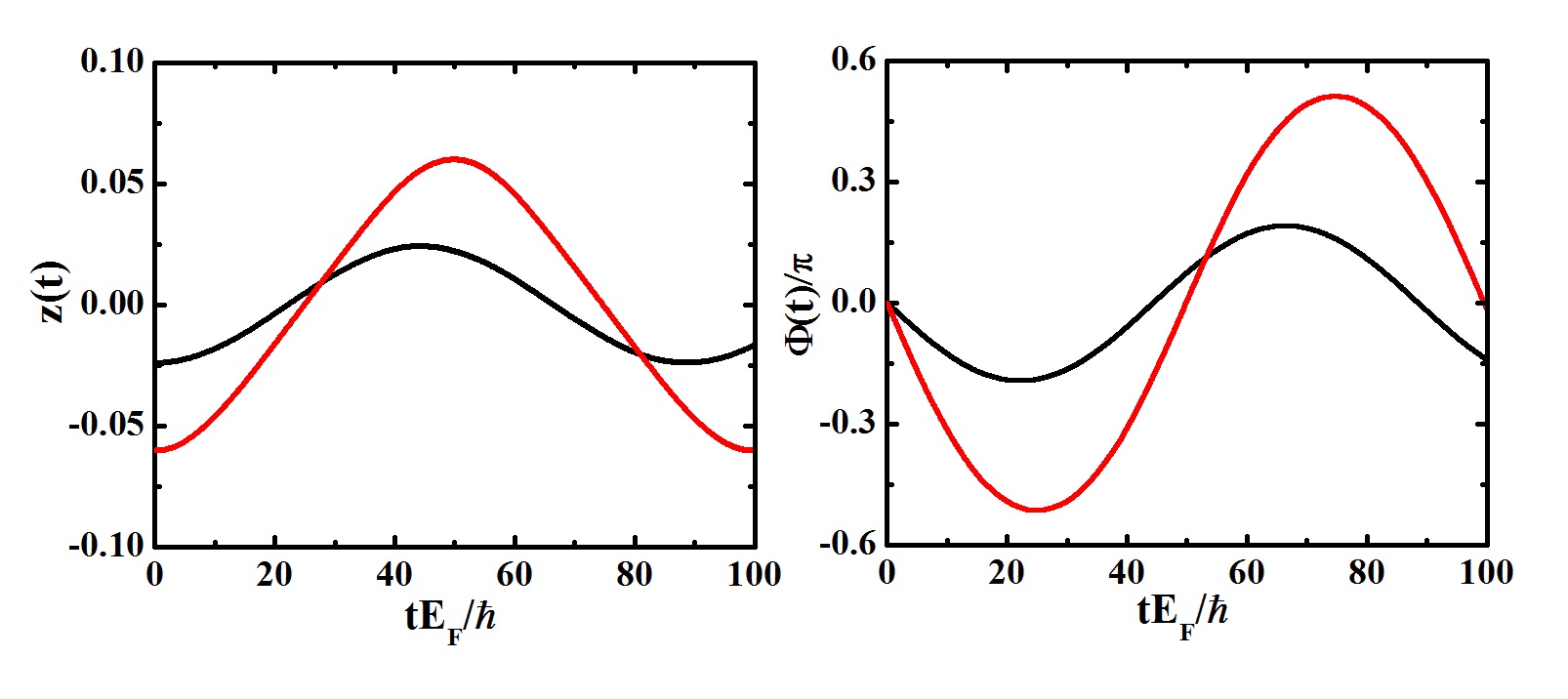}
\caption{Relative population imbalance (left) and phase difference (right) as a function
of time for the same simulation of Fig.~\protect\ref{figjosephson0} ($|z_0|=0.06$, red lines)
and for another simulation with an even smaller initial imbalance ($|z_0|=0.024$, black lines).}
\label{figjosephson1}
\end{figure}

If the initial imbalance is small ($|z_0|\ll 1$), the time evolution of the density and
the order parameter shows clean periodic oscillations. As an example, in
Fig.~\ref{figjosephson0} we show the behavior of the density distribution for an
initial imbalance $|z_0|=0.06$ and with a thin and large barrier ($d=0.6 k_F^{-1}$
and $V_0=5 E_F$).  The evolution of the relative population imbalance
$z(t)$ and the phase difference $\Phi(t)$ is reported in Fig.~\ref{figjosephson1}.
The results for an even smaller imbalance are also shown in the same figure.

\begin{figure}
\includegraphics[scale=0.3]{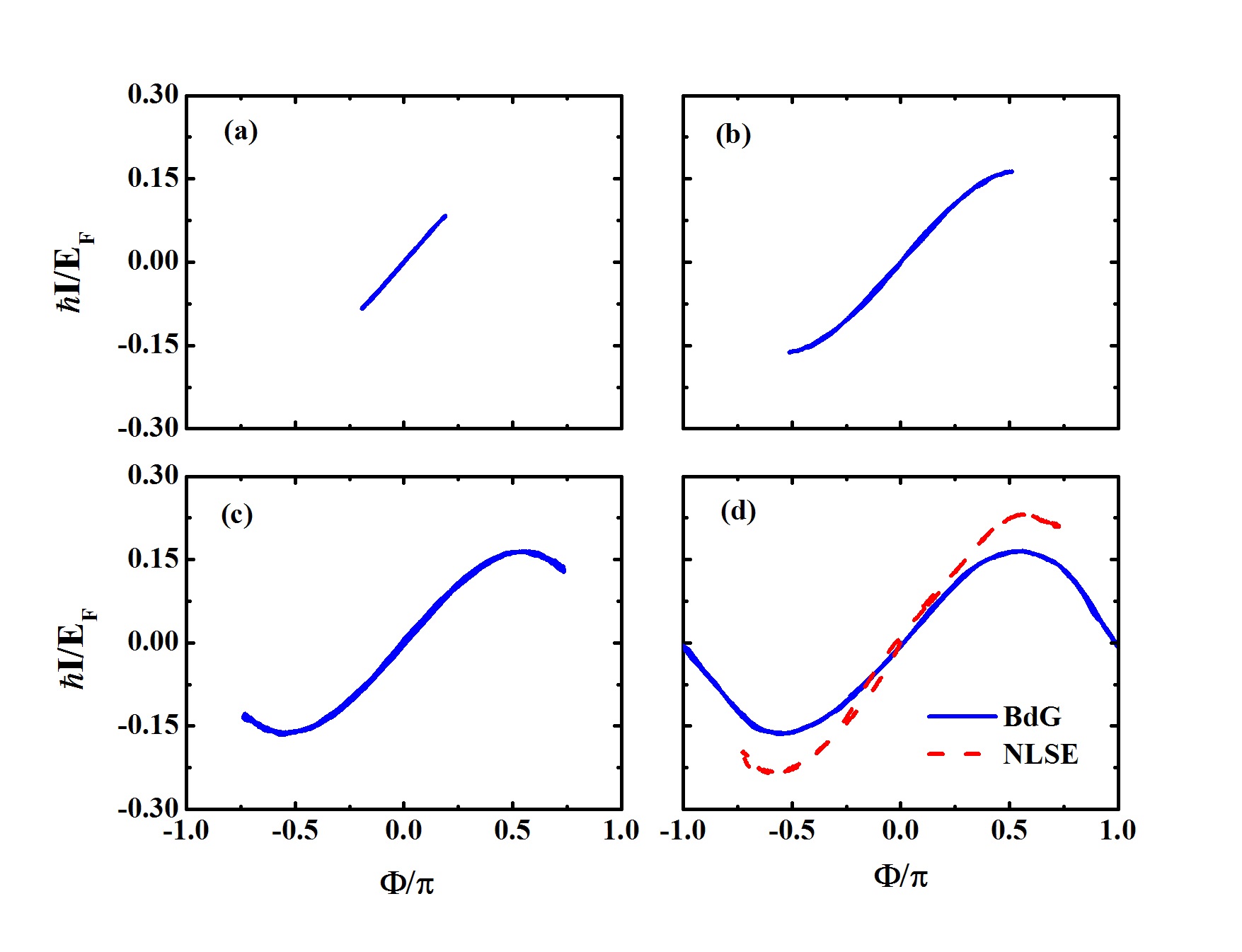}
\caption{The current-phase relation obtained in four BdG simulations with
the same barrier ($V_{0}=5E_F$ and $d=0.6k_F^{-1}$) and different
initial imbalance, $z_0=-0.024$ (a), $-0.06$ (b), $-0.078$ (c), and $-0.096$ (d).  The red
dashed line in panel (d) is obtained by solving the nonlinear Schr\"odinger equation
(\ref{eq:nlse}) in the same configuration and for the same initial imbalance. }
\label{curv5d06}
\end{figure}

Josephson oscillations between weakly linked superfluids ($V_0 \gg \mu$) are
characterized by the sinusoidal relation between current and phase difference \cite{book}:
\begin{equation}
I(t)=I_J \sin \Phi(t)  \;  ,
\label{eq:Jcurrent}
\end{equation}
where the quantity $I_J$ has the meaning of critical Josephson current. In our
case, the current flowing at the barrier position can be easily  calculated as
$I=dN_R/dt=-dN_L/dt=-(N/2) dz/dt$. Fig.~\ref{curv5d06} shows four examples
of the current-phase relation obtained in our simulations with different values
of the initial imbalance. The upper plots correspond to the simulation of
Fig.~\ref{figjosephson1}.

\begin{figure}
\includegraphics[scale=0.3]{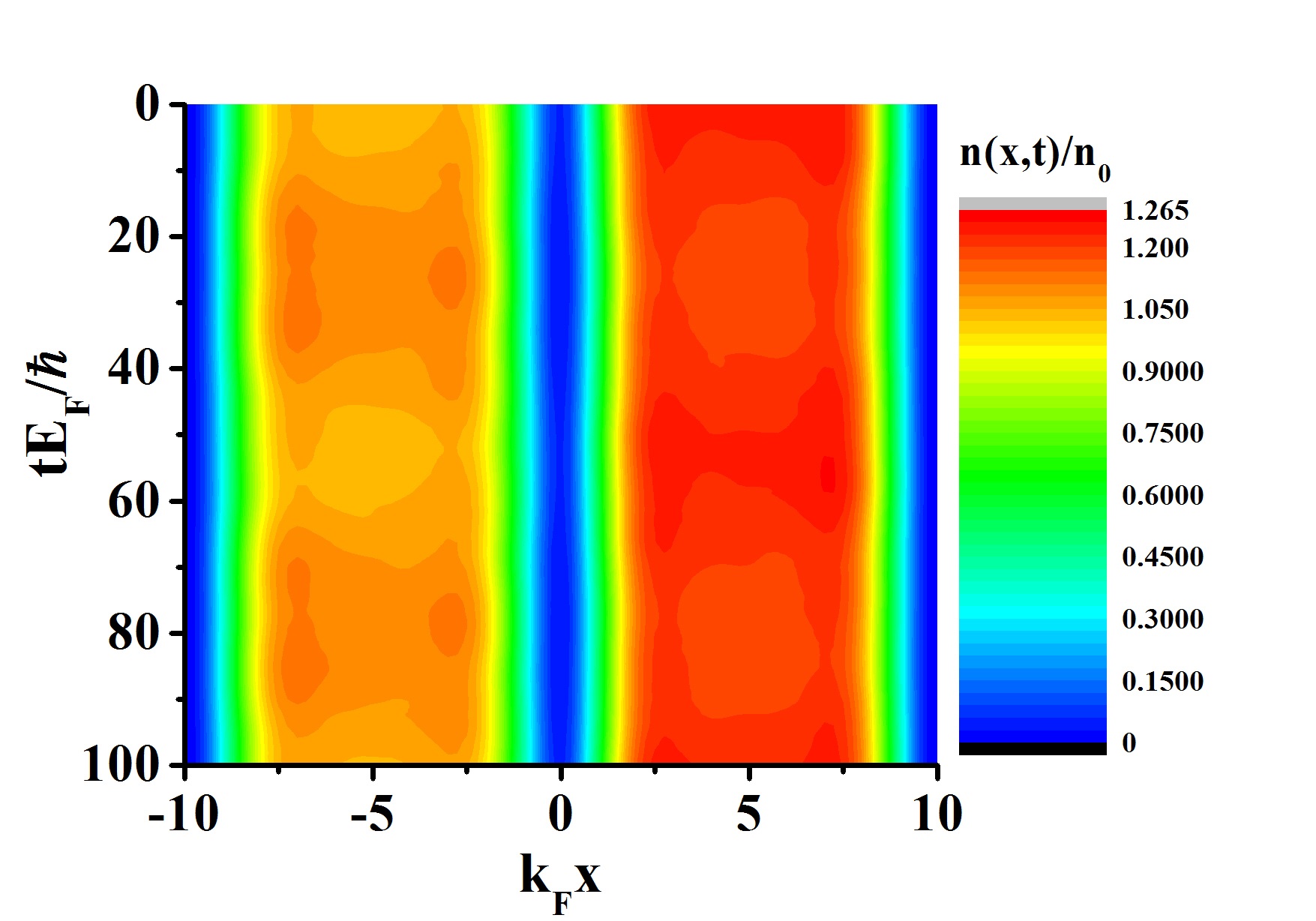}
\caption{Same as in Fig.~\protect\ref{figjosephson0} but for a larger initial
imbalance ($z_0=-0.096$), such to cause self-trapping. }
\label{figselftrapping0}
\end{figure}

\begin{figure}
\includegraphics[scale=0.25]{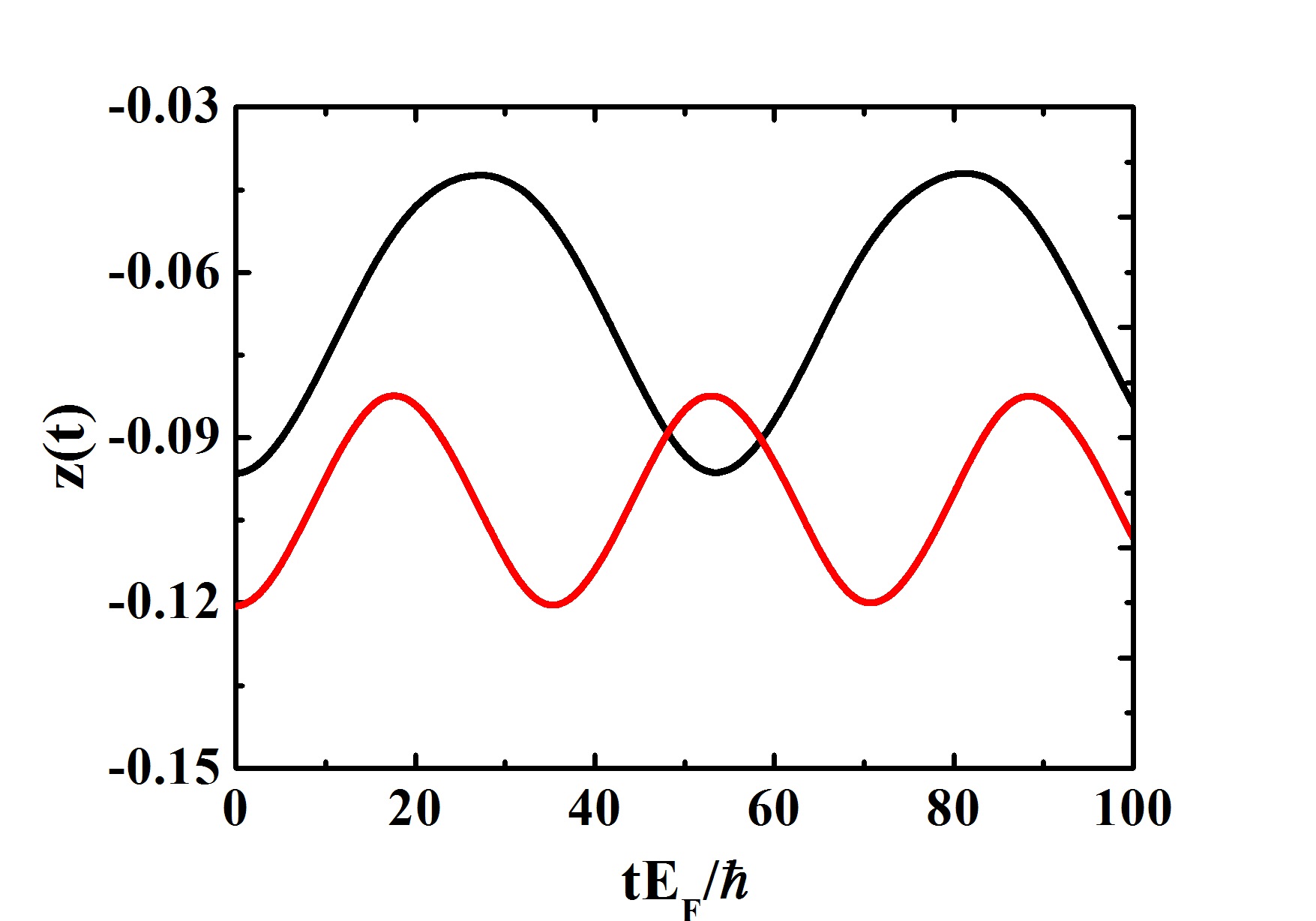}
\caption{Relative population imbalance $z(t)$ for the same
simulation of Fig.~\protect\ref{figselftrapping0}
($|z_0|=0.096$, black line) and another simulation with an even larger
initial imbalance ($|z_0|=0.121$, red line).}
\label{selftrapping1}
\end{figure}

\begin{figure}
\includegraphics[scale=0.3]{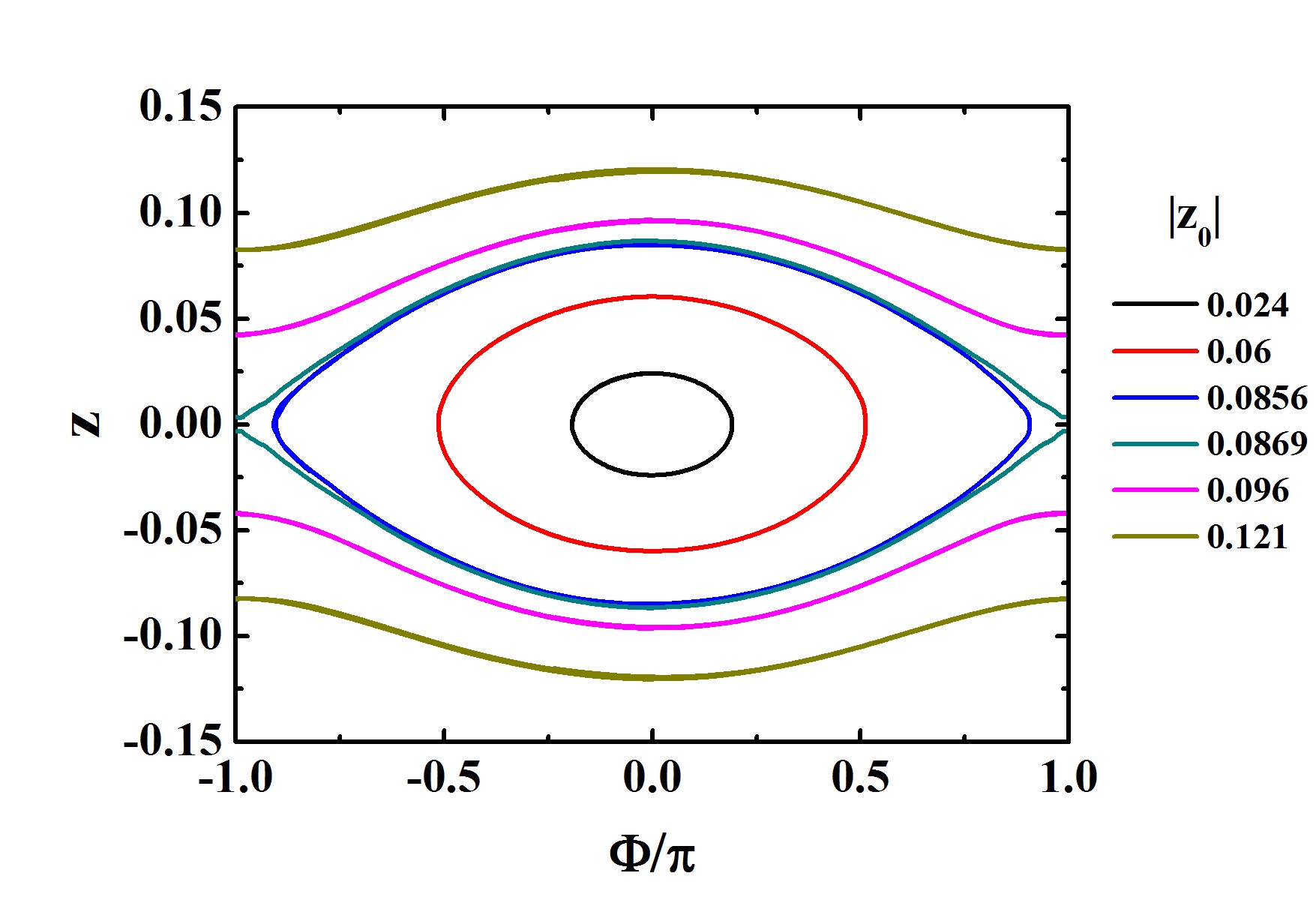}
\caption{Population imbalance {\it vs.} phase difference in simulations with
the same barrier ($V_{0}=5E_F$ and $d=0.6k_F^{-1}$) and different
initial imbalance, $|z_0|=0.024$, $0.06$, $0.0856$, $0.0869$, $0.096$, $0.121$, from the inner
ellipse to the outer open trajectory. The red ellipse corresponds to the simulation
in Fig.~\protect\ref{figjosephson0}; the pink open trajectory corresponds to the
simulation in Fig.~\ref{figselftrapping0}. The transition from Josephson
oscillations to self-trapping occurs at about $|z_0| \approx 0.0869$.  }
\label{phadiav5d06}
\end{figure}

If the initial imbalance exceeds a critical value, the system enters into a
different dynamical regime, where one of the two wells remains always
more populated than the other. The two numbers $N_L$ and $N_R$
oscillate in time, but around unequal mean values. This phenomenon is
known as macroscopic quantum self-trapping \cite{smerzi}. In
Fig.~\ref{figselftrapping0} and \ref{selftrapping1}  we show a typical example.
The transition from the regime of Josephson oscillations and the regime of
self-trapping can be visualized by plotting the trajectories in the diagram of the
population imbalance {\it vs.} the phase difference. Our results for the
barrier with $V_{0}=5E_F$ and $d=0.6k_F^{-1}$ are shown in
Fig.~\ref{phadiav5d06}. Josephson oscillations correspond to close
trajectories, which become elliptic for small amplitudes, while
self-trapping correspond to open trajectories. For the barrier used in
these simulations, the transition between the two regime occurs at
an initial relative imbalance $|z_0|  \approx 0.0869$.

\section{Two-mode model for small Josephson oscillations}

The purpose of this section is to show that the above BdG results for small
oscillations are well reproduced by Josephson junction equations for the two
dynamical variables $z(t)$ and $\Phi(t)$, provided the barrier is large
enough to remain in the weak link regime. In such a situation, the system can be
described as composed by two superfluids located in each well and weakly coupled
by tunneling (two-mode model). Unfortunately, a rigourous derivation of the Josephson
equations from the BdG equations (\ref{tdbdg}) within a two-mode approximation
is not available. We thus proceed by analogy with the case of bosons where, in the Josephson
regime, the population imbalance and the phase difference can be seen as canonically
conjugates variables entering a classical Josephson Hamiltonian of the form \cite{book}
\begin{equation}
H_J=\frac{E_C}{2} k^2 - E_J \cos \Phi \;.
\label{classicalJos}
\end{equation}
The quantity $k$ is defined as $k=(N_L^{\rm (B)}-N^{\rm (B)}_R)/2$, where $N_L^{\rm (B)}$ and $N^{\rm (B)}$
are the number of bosons on the left and right side of the barrrier,  and is assumed to
be small. The quantities $E_C$ and $E_J$ have the meaning of on-site energy  (local
interaction within each well) and tunneling energy (or Josephson coupling energy), respectively.
 From (\ref{classicalJos}) one gets the equations of motion
\begin{eqnarray}
\frac{\partial k}{\partial t} & = &  -\frac{\partial H_J}{\partial (\hbar \Phi)}
 =   - \frac{E_J}{\hbar} \sin \Phi   \label{dotk}  \\
\frac{\partial \Phi}{\partial t} & = &
\frac{\partial H_J}{\partial (\hbar  k)}  =   \frac{E_C}{\hbar} k  \label{dotphi} \; .
\end{eqnarray}
If $|\Phi| \ll 1$, the two equations admit harmonic solutions
corresponding to Josephson oscillations of frequency
\begin{equation}
\omega_p = \frac{1}{\hbar} \sqrt{E_C E_J}
\label{periodplasma}
\end{equation}
also known as plasma frequency. These results are valid in the
Josephson regime where $E_C/E_J$ is of order $1$ or less, but much larger than
$N^{-2}$; different regimes are obtained when $E_C/E_J \ll N^{-2}$ (Rabi
regime) and $E_C/E_J \gg 1$ (Fock regime) \cite{leggett,gati,smerzi}.

In order to check the applicability of this scheme to the BdG results of the previous
section, we need to know how to calculate $E_C$ and $E_J$ within the same theory.
We first notice that the tunneling energy $E_J$ can be easily related to the energy
difference $\Delta E= E^--E^+$, where  $E^+$ and $E^-$ are the energies of
the lowest symmetric and antisymmetric states in the double-well potential with
zero imbalance ($k=0$). In fact, these states have $\Phi=0$ and $\Phi=\pi$,
respectively, and hence the Hamiltonian (\ref{classicalJos}) gives $E_J=\Delta E/2$.
Moreover we can relate both $E_J$ and $\Delta E$ to the Josephson current $I_J$.
In fact, the number of bosons, i.e., pairs of fermionic atoms, tunneling through
the barrier at $x=0$ per unit time is $I^{\rm (B)}=-dk/dt$, so that the current
of atoms is $I=2I^{\rm (B)}= (2E_J/\hbar) \sin \Phi$, as in Eq.~(\ref{eq:Jcurrent}),
with $I_J = 2E_J/\hbar=\Delta E/\hbar$.

A nice feature of the last relation is that it can be numerically tested by performing two
independent calculations. On one hand,  the Josephson current
$I_J$ can be obtained by solving the time dependent BdG equations (\ref{tdbdg}):
by looking at the current-phase plots, like those in Fig.~\ref{curv5d06},  the
current $I_J$ can be extracted as the maximum of the curve. On the
other hand,  the energy difference $\Delta E$ can be calculated
by solving the stationary BdG equations (\ref{statBdG})
for the ground (symmetric) state and the lowest antisymmetric state (see details in the
Appendix).  In Fig.~\ref{IJdJ}  we show the results obtained with $100$ particles in a
box of size $L=20k_F^{-1}$ and different barriers. The figure
shows that the relation $I_J =\Delta E/\hbar$ is remarkably well satisfied.

\begin{figure}
\includegraphics[scale=0.25]{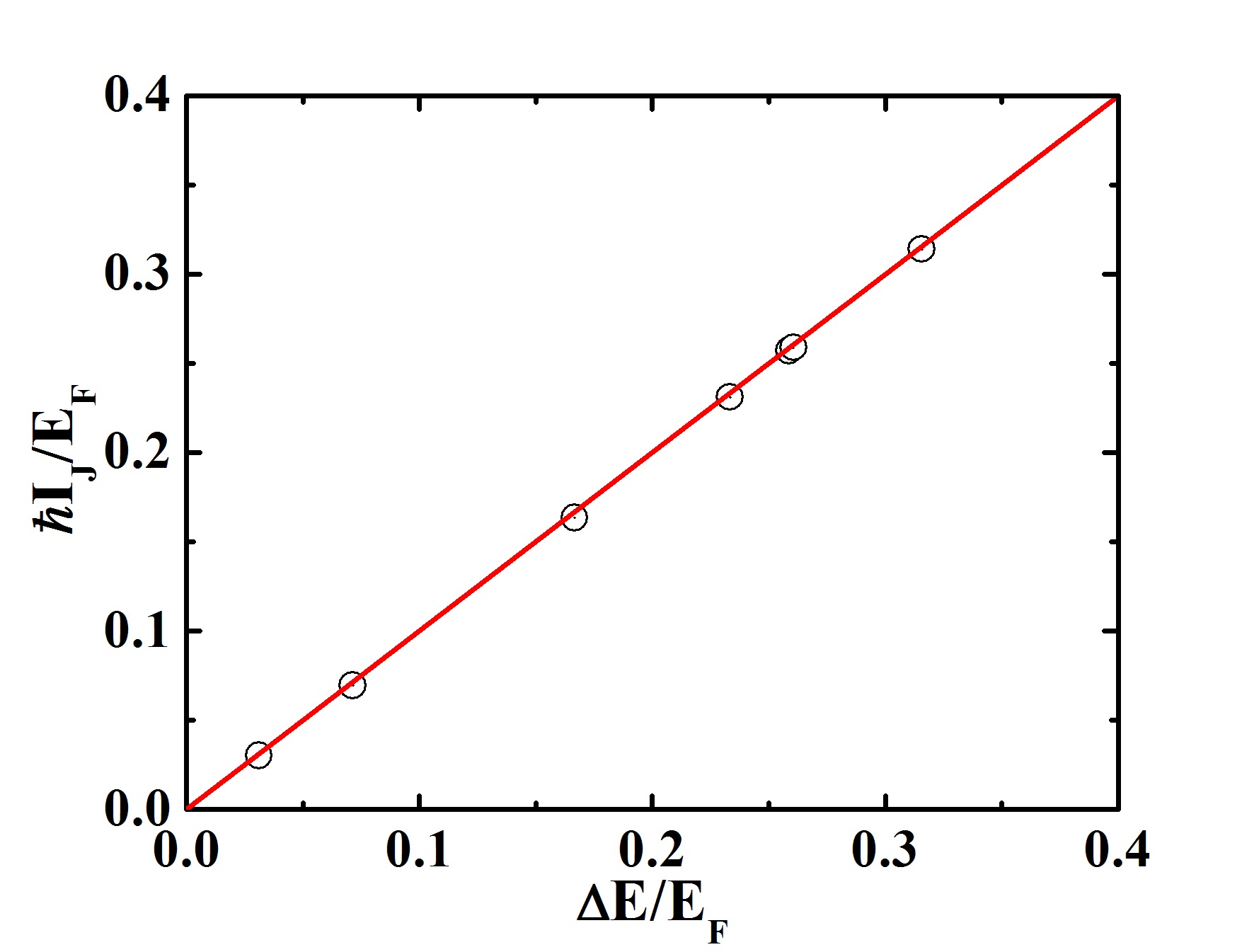}
\caption{The Josephson current $I_J$ extracted from time dependent BdG simulations
in the regime of small oscillations and weak tunneling is plotted as a function of the
energy difference $\Delta E= E^--E^+$, between  the
lowest antisymmetric and symmetric solutions of the stationary BdG  equations (\ref{statBdG}).
All points correspond to $E_{\rm cut}= 50 E_F$, $N=100$, $L=20k_F^{-1}$ and
$L_\perp=13k_F^{-1}$, while ($V_0/E_F$, $k_Fd$) is (5,1), (5,0.8), (5,0.6), (6,0.45), (5,0.5), (4,0.6), (4,0.55),
for points from bottom-left to top-right. The red line represents the equality
$\hbar I_J=\Delta E$.
}
\label{IJdJ}
\end{figure}

\begin{figure}
\includegraphics[scale=0.25]{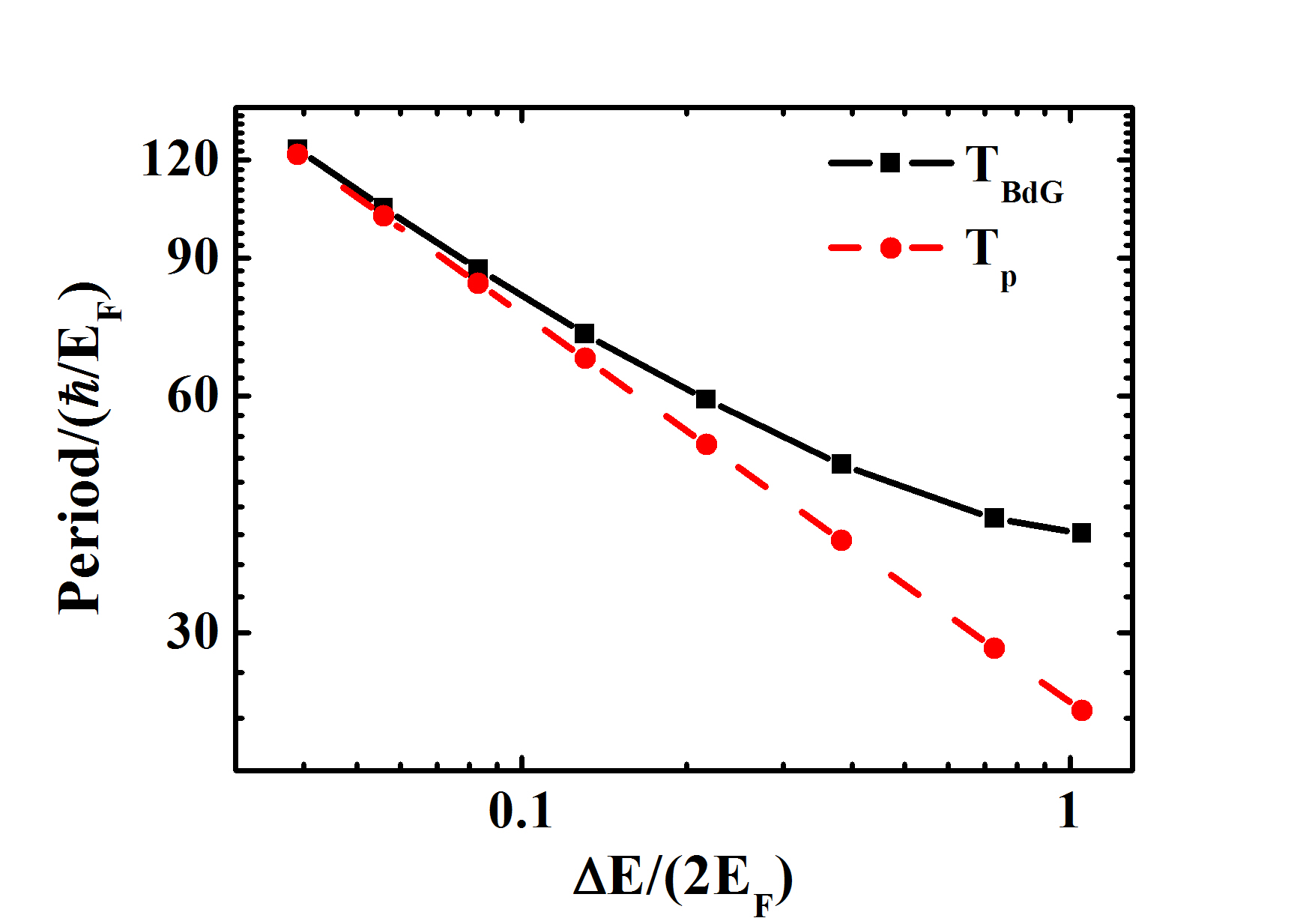}
\caption{Period of small amplitude Josephson oscillations as a function of
$\Delta E/2$, in log-log scale. The black solid line is the period $T_{BdG}$
observed in time dependent BdG simulations for $100$ atoms in a double-well
potential. For each black point the width of the barrier is the same, $d=0.6k_F^{-1}$,
while its height decreases from $V_0=7 E_F$ (leftmost point) to $0.6 E_F$ (rightmost
point).  The red dashed line is the period  $T_p = 2\pi/\omega_p
= h /\sqrt{E_C E_J}$ of plasma oscillations, where $E_C$ and $E_J$ are calculated
by solving the stationary BdG equations (\ref{statBdG}),  assuming $E_J=\Delta E/2$. }
\label{period}
\end{figure}

The on-site energy $E_C$ accounts for the variation of the interaction energy of the
system due to the exchange of particles between the two wells. For a bosonic
superfuid in a symmetric well, this parameter is given $E_C = 2 d\mu^{\rm (B)}/dN_L^{\rm (B)}$
\cite{book},  where $\mu^{\rm (B)}$ is the chemical potential and its derivative is calculated
at $N_L^{\rm (B)}=N^{\rm (B)}/2$. Expressing the same quantity in terms of the
chemical potential of the fermionic atoms and the number of atoms, we can write
$E_C = 8 d\mu/dN_L$. This quantity can be obtained by solving the
stationary BdG equations (\ref{statBdG}) for different atom numbers
in the same double-well. Having $E_J$ and $E_C$, we can finally calculate
the plasma period $T_p = 2\pi/\omega_p = h /\sqrt{E_C E_J}$
and compare it with the period of the oscillations observed in the time-dependent
BdG simulations. The comparison is reported in Fig.~\ref{period}, where we plot
$T_{\rm BdG}$ (black
solid line) and $T_p$ (red dashed line) as a function of $E_J=\Delta E/2$. As one can see,
in the limit of small tunneling ($\Delta E \to 0$), the period observed in the
BdG simulations nicely approach the plasma period $T_p$.

These results show that the small oscillations of  two weakly coupled fermionic
superfluids at unitarity, as obtained with the BdG equations, can be  accurately
reproduced by a two-mode model for Josephson oscillations.

\section{Large oscillations and self-trapping}

Let us now consider larger oscillations and the transition to self-trapping. The
classical Josephson Hamiltonian (\ref{classicalJos}) does not apply anymore and
we may wonder whether nonlinear effects can be properly included in a two-mode
model. For Bose-Einstein condensates governed by the Gross-Pitaevskii equation,
coupled nonlinear Josephson junction equations for the number imbalance and the
phase were analytically derived by  Smerzi {\it et al.} \cite{smerzi}. A similar derivation
is also available for fermions in the BCS-BEC crossover within a phenomenological
density functional theory \cite{salasnich2}. This theory is based on the use of the
following nonlinear Schr\"odinger equation (also named density functional GP equation,
or extended Thomas-Fermi equation)
\begin{equation}
i\hbar\frac{\partial}{\partial t}\Psi({\bf r},t) = [-\frac{\hbar^2}{4m}\nabla^2+2V_{ext}({\bf r})
+ \mu_{\rm loc}^{(B)}(n,a)]\Psi({\bf r},t)
\label{eq:nlse}
\end{equation}
for the order parameter $\Psi$ of Cooper pairs of mass $2m$, with
$|\Psi({\bf r})|^2= n({\bf r})/2$, if $n$ is the atom density. The key ingredient of this
nonlinear Schr\"odinger equation is the local "bulk" chemical potential of Cooper
pairs, $\mu_{\rm loc}^{(B)}(n,a)=2\mu(n,a)+\hbar^2/(ma^2)$, where
$\mu(n,a)$ is the chemical potential of a uniform Fermi gas of density $n$ and the
second term is the binding energy of the pair. Its expression is an input of the theory; it
can be taken from ab initio Monte Carlo calculations of the equation of state or from
the mean-field BdG theory, or different suitable parametrizations. Once
$\mu_{\rm loc}(n,a)$ is given, the NLSE  (\ref{eq:nlse}) can be numerically
solved for studying stationary and/or time dependent configurations.
The advantages and the limits of this approach have been widely discussed in
the literature (see for instance the recent discussion in \cite{forbes}, and references
therein). Here we only focus on the fact that, when applied to a double-well
potential in the weak link limit, the NLSE can be cast into the form of Josephson
junction equations \cite{salasnich2}. This is done by assuming the order parameter
to be a superposition of the left and right parts,
\begin{equation}
\Psi({\bf r},t)=c_L(t)\Psi_L({\bf r})+c_R(t)\Psi_R({\bf r})
\label{2mode}
\end{equation}
having an exponentially small overlap under the central barrier. By inserting this
ansatz for $\Psi$ into Eq.~(\ref{eq:nlse}), after integration over space and neglecting
exponentially small $\Psi_L\Psi_R$ terms, one obtains the equations
\begin{eqnarray}
i\hbar\frac{\partial}{\partial t} c_L(t) & = & E_Lc_L(t) - {\cal K} c_R(t)
\label{2mode1} \\
i\hbar\frac{\partial}{\partial t} c_R(t) & = & E_Rc_R(t) - {\cal K} c_L(t)
\label{2mode2}
\end{eqnarray}
for the two complex coefficients $c_i(t)$ in region $i$, with $i=L,R$.
The energy $E_i=E_i^0+E_i^I$ is the sum of
\begin{equation}
E^0_i(\sqrt{N_i})=\int d{\bf r}\Psi_i({\bf r})[-\frac{\hbar^2}{4m}\nabla^2+2V_{\rm ext}({\bf r})]\Psi_i({\bf r})
\end{equation}
\begin{equation}
E^I_i(\sqrt{N_i})=\int d{\bf r}\Psi_i({\bf r})\mu^{(B)}_{\rm loc}(n_i,a)\Psi_i({\bf r}) \; ,
\end{equation}
while the coupling term is given by
\begin{equation}
{\cal K}= - \int d{\bf r}\Psi_L({\bf r})[-\frac{\hbar^2}{4m}\nabla^2+2V_{\rm ext}({\bf r})]\Psi_R({\bf r}) \;.
\end{equation}
The functions $\Psi_R(\textbf{r})$ and $\Psi_L(\textbf{r})$ are real, obey the
orthonormality condition $\int d \textbf{r} \Psi_i \Psi_j = \delta_{i,j}$ and are localized in each of
the two wells.  In a symmetric system (i.e., $V_{\rm ext}(-\rm \textbf{r})=V_{\rm ext}(\rm \textbf{r})$), one
has $\Psi_R(-\textbf{r})=\Psi_L(\textbf{r})$ and thus $E_L^0=E_R^0$ and
$E_L^I=E_R^I=E^I$. By writing $c_{L,R}=\sqrt{N_{L,R}/2}
\exp (i\phi_{L,R})$ and inserting it into Eqs.~(\ref{2mode1}) and (\ref{2mode2}),
one gets \cite{salasnich2}
\begin{eqnarray}
\frac{\partial z}{\partial t} & = & -\frac{2{\cal K}}{\hbar}\sqrt{1-z^{2}}\sin\Phi
\label{eq:AD_dn} \\
\frac{\partial\Phi}{\partial t}
 &  = & \frac{1}{\hbar}[E^I(\sqrt{N_L})-E^I(\sqrt{N_R})]
+\frac{2{\cal K}}{\hbar}\frac{z\cos\Phi}{\sqrt{1-z^{2}}}
\label{eq:ge_phase}
\end{eqnarray}
where the imbalance $z$ and the phase difference $\Phi$ are the same already
defined at the beginning of section III.

\begin{figure}
\includegraphics[scale=0.3]{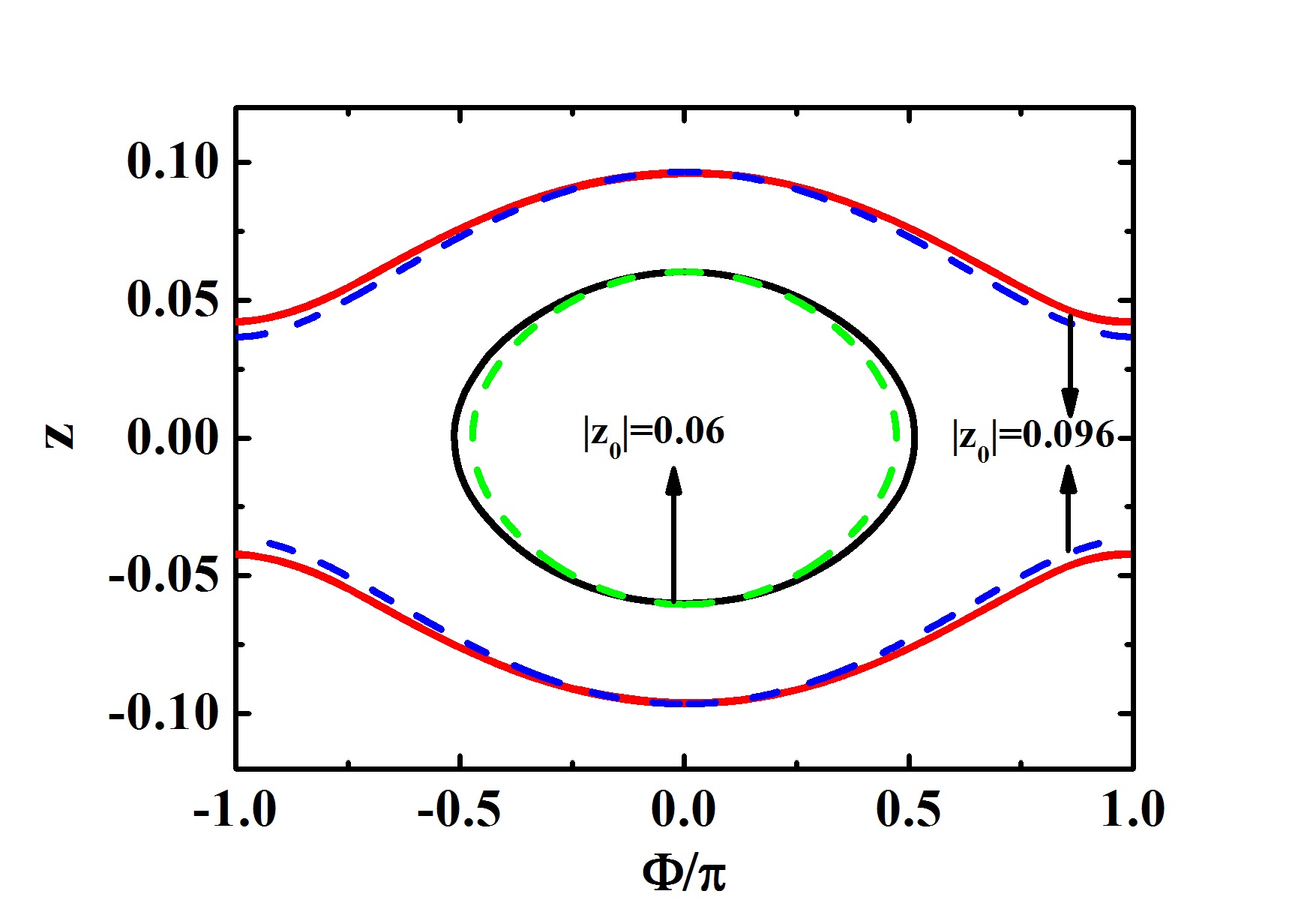}
\caption{Population imbalance {\it vs.} phase difference. The potential barrier has
height $V_{0}=5E_F$ and width $d=0.6k_F^{-1}$, as in Fig.~\protect\ref{phadiav5d06}.
Solid lines are the results of time-dependent BdG simulations, while
dashed lines are the solutions of the  nonlinear Josephson equations (\ref{eq:AD_dn})
and (\ref{eq:AD_phase}), with $E_J=0.0833E_F$ and $E_C=0.0678E_F$ taken from the
solutions of the stationary BdG equations. Closed trajectories correspond to Josephson
oscillations with initial imbalance $|z_0|=0.06$, while open trajectories correspond to
self-trapped states with $|z_0|=0.096$. }
\label{comV5}
\end{figure}

At unitarity the chemical potential of the uniform Fermi gas of density $n$ is
$\mu(n)=(1+\beta)E_F(n)$,  where $\beta$ is a universal parameter \cite{rmp08}. This implies
$E^I(\sqrt{N_i})=U(N_i/2)^{2/3}$  with $U=[\hbar^2 (3\pi^2)^{2/3}(1+\beta)/m]
\int\! d{\bf r}\Psi_i^{10/3}$, and Eq.~(\ref{eq:ge_phase}) becomes
\begin{equation}
\frac{\partial\Phi}{\partial t}
=\frac{2{\cal K}}{ \hbar} \left( \Lambda[(1+z)^{2/3}-(1-z)^{2/3}]
+\frac{z\cos\Phi}{\sqrt{1-z^{2}}}  \right)
\label{eq:AD_phase}
\end{equation}
where $\Lambda=(N/4)^{2/3}U/2{\cal K}$ \cite{adhikari,noteN}.
The corresponding classical Hamiltonian is
\begin{equation}
\frac{H}{2{\cal K}}=\frac{3\Lambda}{5}[(1+z)^{5/3}+(1-z)^{5/3}]-\sqrt{1-z^2}\cos\Phi \; .
\label{eq:Hclassic}
\end{equation}
In the limit of small amplitude oscillations ($|\Phi| \ll 1$ and $|z| \ll 1$),
the equations of motion (\ref{eq:AD_dn}) and (\ref{eq:AD_phase}) reduce to the
linear Josephson equations (\ref{dotk}) and (\ref{dotphi}) provided the two parameters
$\Lambda$ and ${\cal K}$ are related to the on-site interaction energy $E_C$ and the
tunneling energy $E_J$ by
\begin{equation}
{\cal K}=\frac{2E_{J}}{N} \;  , \; \Lambda=\frac{3}{4} \left( \frac{N^{2}E_{c}}{16E_{J}}-1 \right) \; .
\label{eq:kappalambda}
\end{equation}

At this point we are ready to compare our BdG results of section III with the
two-mode model including the nonlinear regime. For each configuration (i.e., for each
set of parameters $L, L_\perp , V_0, d, N$) we can calculate the two energies
$E_C$ and $E_J$ by solving the stationary BdG equations as explained in section IV.
Then we can use them in (\ref{eq:kappalambda}) to calculate ${\cal K}$ and $\Lambda$ and
solve the nonlinear Josephson equations (\ref{eq:AD_dn}) and (\ref{eq:AD_phase}) for
different values of the initial population imbalance. The results can then be compared
with those obtained by solving the time dependent BdG equations (\ref{tdbdg}).
In Fig.~\ref{comV5} we show typical results for the imbalance vs. phase diagram, for the
same configuration of Fig.~\ref{phadiav5d06}. The  agreement between BdG equations
(solid lines) and nonlinear Josephson equations (dashed lines) is remarkably good both
in the case of Josephson oscillations (inner ellipse) and self-trapping (open trajectories).
In the BdG simulations the transition between the two regimes occurs at $|z_0|
\approx 0.0869$. In the case of the nonlinear Josephson equations (\ref{eq:AD_dn})
and (\ref{eq:AD_phase}) the same transition is obtained when the energy (\ref{eq:Hclassic})
reaches the critical value \cite{adhikari}
\begin{equation}
E_{\rm cr}=2{\cal K} \left( \frac{6\Lambda}{5}+1 \right) =
\frac{4E_J}{N} \left( \frac{9N^2E_C}{160E_J}+\frac{1}{10} \right) \; .
\label{eq:ecr}
\end{equation}
For the parameters of Fig.~\ref{comV5}, this condition corresponds to $|z_0| \approx 0.0893$,
which is again very close to the BdG result.

The agreement between BdG equations and nonlinear Josephson equations is not restricted
to unitarity. We tested that a similar agreement is found also for $1/(k_Fa) \neq 0$, both at the
BEC side ($1/(k_Fa) > 0$) and BCS side ($1/(k_Fa) < 0$)
of the BCS-BEC crossover. This suggests that the validity of the nonlinear
Josephson equations (\ref{eq:AD_dn}) and (\ref{eq:ge_phase})  is more general than
the validity of the NLSE (\ref{eq:nlse}) which is known to be accurate in the
BEC regime but not in the BCS regime, where it misses the fermionic degrees of freedom.
In Ref.~\cite{salasnich2} it was noticed that, despite this inaccuracy of the NLSE,
the nonlinear Josephson equations can still be used in the whole crossover, provided the
tunnelling energy is taken as a phenomenological parameter. Our numerical results show
that the same nonlinear Josephson equations are a very good approximation of the weak link
limit of the BdG equations, the parameters $E_C$ and $E_J$ being consistently calculated
within the same BdG theory.

\begin{figure}
\includegraphics[scale=0.3]{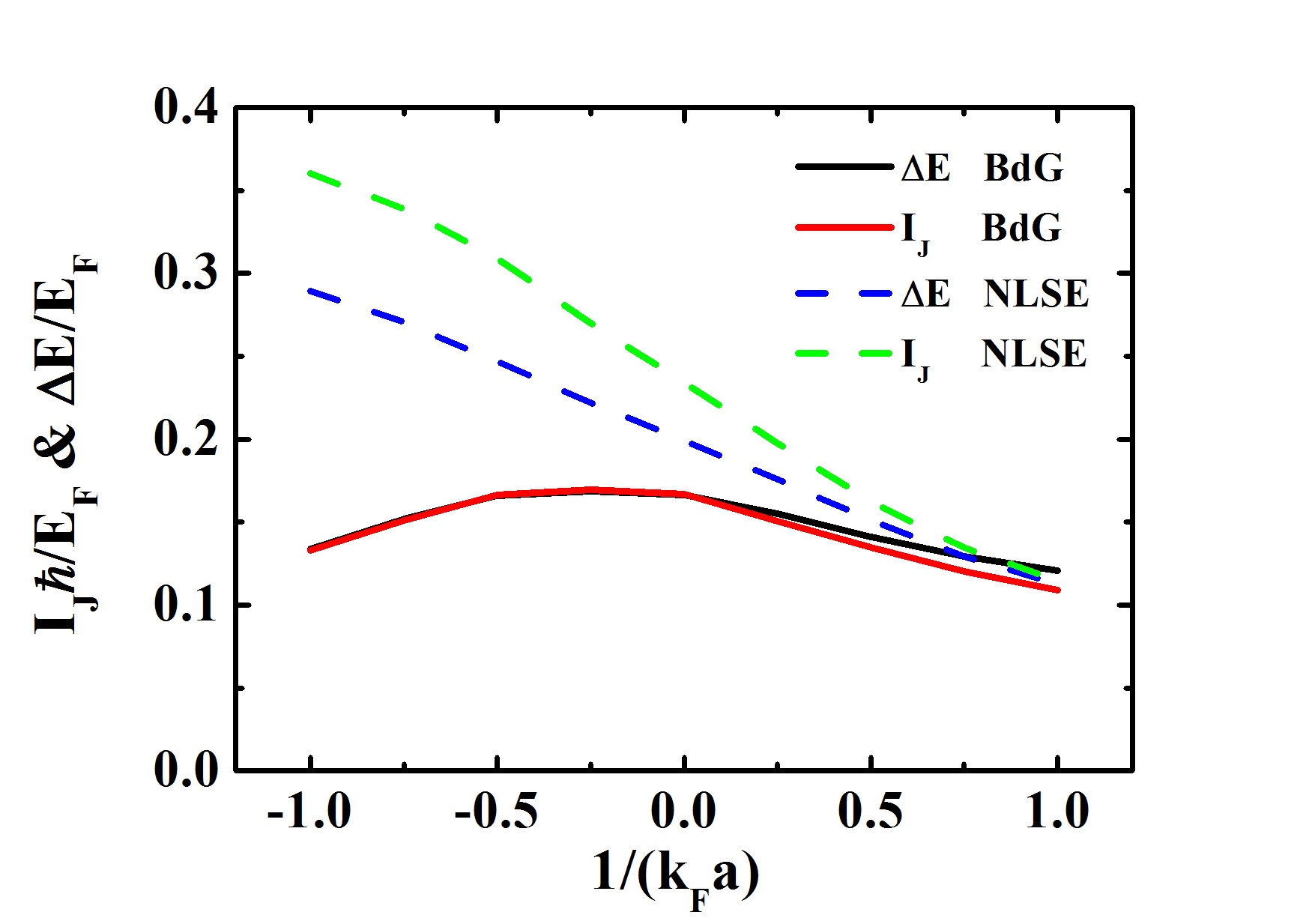}
\caption{Energy difference $\Delta E$ and maximum Josephson current $I_J$
calculated with the BdG equations (solid lines) and the NLSE (dashed lines), as a
function of the interaction strength $1/(k_Fa)$. The
parameters of the barrier are $d=0.6 k_F^{-1}$ and $V_0=5 E_F$, and the
number of atoms $N=100$.  }
\label{nlseVSBdGEJIJ}
\end{figure}

The difference between NLSE and BdG equations can be appreciated by looking at
Fig.~\ref{nlseVSBdGEJIJ}, where we plot the results for the maximum Josephson current,
$I_J$, together with the energy difference $\Delta E$. The quantity $I_J$ is extracted
from time dependent simulations, either solving the BdG equations (\ref{tdbdg}) (red
solid line) or the NLSE (\ref{eq:nlse}) (upper dashed line), while $\Delta E$ is calculated
from the corresponding stationary (time independent) equations; in Eq.~(\ref{eq:nlse})
we use the mean-field equation of state (MF EOS) for the local chemical potential
\cite{salasnich2}. As discussed in section IV, in the weak link limit, where the nonlinear
Josephson equations are expected to hold, the quantity $\Delta E$ should be equal to
twice the tunneling energy $E_J$ and one should find $\hbar I_J = \Delta E$. This is
clearly the case for BdG equations where $\Delta E$ (black solid line) and $\hbar I_J$
(red solid line) are almost indistinguishable in the whole crossover,
the small difference in the BEC limit being likely due to the finite cutoff energy in the
BdG calculations, which becomes a more critical parameter as $1/(k_Fa)$ increases.
Conversely in the case of NLSE, the two quantity are significantly different and the
critical current $I_J$ is increasingly larger than the BdG prediction in the BCS limit.
The difference can be seen also in Fig.~\ref{curv5d06} where we show an example
of Josephson oscillations at unitarity as obtained by solving Eq.~(\ref{eq:nlse})
(dashed line) and Eq.~(\ref{tdbdg}) (solid line) for the same configuration.
The fact that  $I_J$ is larger in the NLSE than in BdG equations is well known and
is simply due to pair-breaking processes which are included in BdG  \cite{combescot}
but are absent in the NLSE. This effect was already discussed in Ref.~\cite{salasnich2} in a regime
of wider ($d > k_F^{-1}$) and lower ($V_0<E_F$) barriers. Here, on purpose, we have
chosen thinner barriers, i.e, $d$ of the order or less than $k_F^{-1}$, in order to test
the applicability of the two-mode model to cases where density and phase variations
occur on the lengthscale of the inverse Fermi wave vector, such that the local density
approximation becomes questionable and fermionic degrees of freedom might play a role.
Our results indicate that, at least in the weak link
limit and within a mean-field theory, the dynamics is still dominated by tunneling of
bosonic pairs and is suprisingly well described by the nonlinear Josephson
equations (\ref{eq:AD_dn})-(\ref{eq:ge_phase}).

\begin{figure}
\includegraphics[scale=0.3]{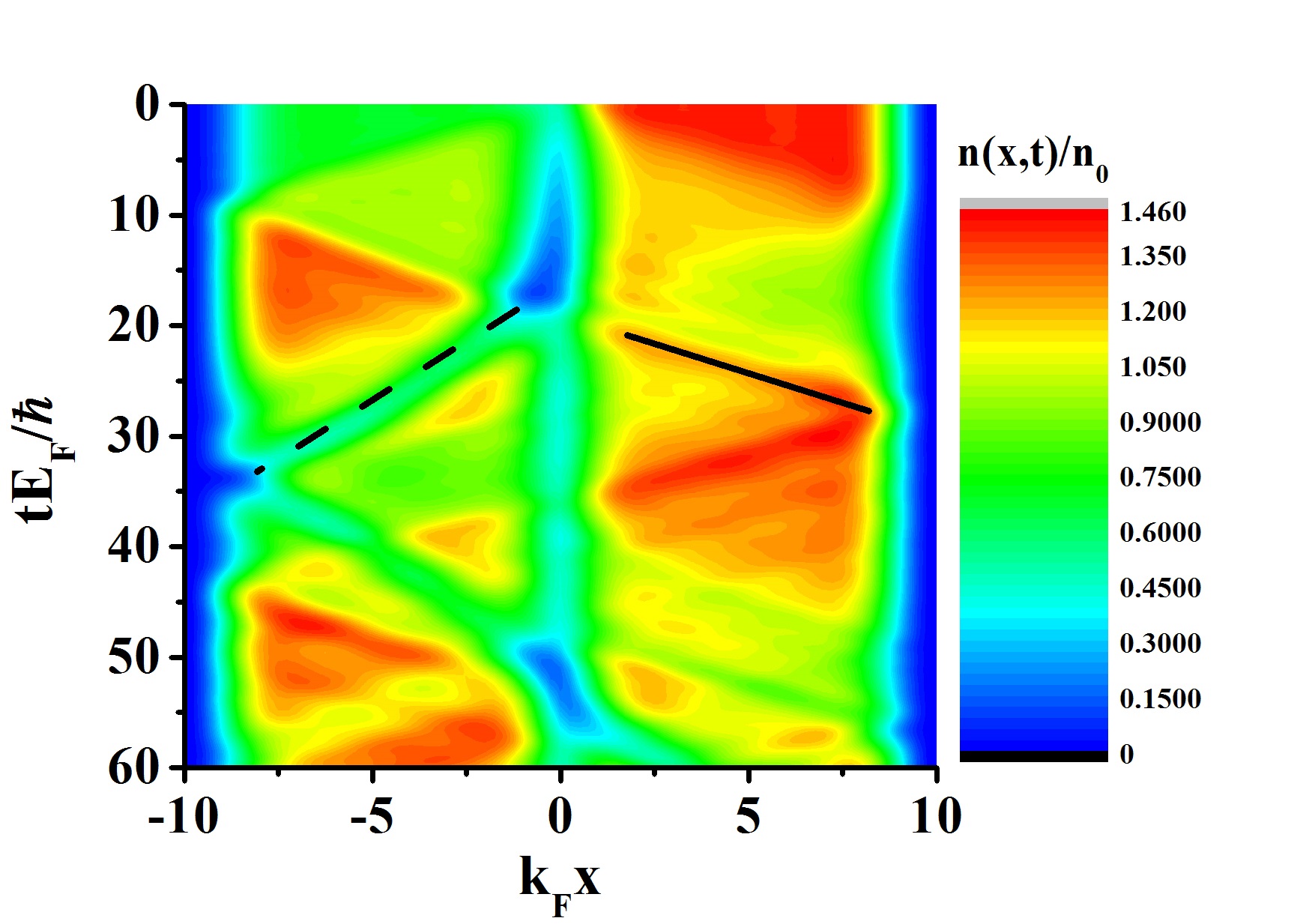}
\caption{Evolution of the density distribution $n(x,t)/n_0$ at unitarity, obtained
by solving the time-dependent BdG equations (\ref{tdbdg}), as in
Fig.~\protect\ref{figjosephson0}, but for a lower barrier ($V_0=E_F$
and $d=0.6k_F^{-1}$). The initial imbalance is $|z_0|=0.353$.
Solid and dashed lines represent the propagation of a sound-like density wave
packet and a grey soliton, respectively. }
\label{v1d06den}
\end{figure}

The situation is rather different when the coupling
between the two wells is strong. An example is shown in Fig.~\ref{v1d06den},
where we plot the density $n(x,t)$ in a BdG simulation with a low barrier
($V_0=E_F$ and $d=0.6k_F^{-1}$) and large initial imbalance. The
Josephson current through the barrier is strongly coupled to the collective motion
of the gas in the two wells. One can distinguish a density wave bouncing
back and forth with a velocity of the order of the
sound speed in a unitary Fermi gas with the same average density, $\sqrt{(1+\beta)/3}
 \ v_F$\cite{rmp08}. In addition, at about $t=15 \hbar/E_F$, when the density
under the barrier almost vanishes, a grey soliton is nucleated. The soliton appears
as a density depletion travelling leftward (dashed line) at a velocity smaller
than the speed of sound. The phase of the order parameter has a variation of
the order of $\pi$ across the soliton. In the case of an infinite system, this
mechanism of soliton nucleation induces a dissipation of the superfluid
current due to phase slip \cite{piazza}. In our confined double-well system,
solitons and collective sound-like waves are coupled by nonlinear mixing and
eventually lead to a decay of the initial Josephson oscillation.

\section{Conclusions}

Two weakly linked Fermi superfluids in a double well geometry exhibit dynamical
regimes of Josephson oscillations and self-trapping. The nonlinear Josephson equations
that describe both regimes are the analog of those for Bose superfluids
described by the GP equation \cite{smerzi}; indeed the fermionic versions of such equations
were previously derived starting from a generalized GP equation \cite{salasnich2}. Here we
show that the same nonlinear Josephson equations are in remarkable agreement with time-dependent
BdG simulations, with the on-site energy, $E_C$, and the tunneling energy, $E_J$, consistently
calculated within the same BdG theory. Such an agreement could not be foretold {\it a priori} since,
on one hand,  a formal derivation of the nonlinear Josephson equations from the
BdG equations is yet missing and, on the other hand, the role played by fermionic
degrees of freedom in the dynamics of a superfluid subject to density and phase variations
on  the lengthscale of the inverse Fermi wave vector is largely unkown. Our results indicate
that the dynamics at the weak link is dominated by tunneling of bosonic pairs, with the caveat
that the critical Josephson current $I_J$ in the BCS side of the crossover is determined by
pair-beaking processes. For lower barriers and large tunneling, the dynamics involves
the collective motion of the superfluid in the two wells and possible phase slip processes due
to the nucleation of solitons. Our predictions complement those of Ref.~\cite{strinati}
and are intended to stimulate new experimental investigations with ultracold Fermi gases in
double-well potentials.

\begin{acknowledgments}
We thank S.~Stringari, L.P.~Pitaevskii, A.~Recati for fruitful discussions. P.Z is
particularly indebted to R.G.Scott, whose computational expertise was of great help
at the beginning of this work.  Support of ERC, through the QGBE grant, and of
Provincia Autonoma di Trento is acknowledged.
\end{acknowledgments}

\section*{Appendix:  Calculation of the energy difference $\Delta E$} \label{sec:appendix}

In this appendix we provide some details about the calculations of the energy
difference, $\Delta E=E^--E^+$,  where $E^+$ is the energy of the (symmetric)
ground state of the gas and $E^-$ is the energy of the lowest anti-symmetric
state with the same number of particles $N$ in the box. Here symmetric and
anti-symmetric refer to spatial reflection in the $x$-direction around $x=0k_F^{-1}$.
The anti-symmetric state corresponds to the solutions of Eq.~(\ref{statBdG})
exhibiting a $\pi$ phase jump in the order parameter $\Delta$ when
crossing the center of the box. For weakly coupled superfluids the quantity
$\Delta E$ is directly related to the Josephson tunnelling energy, $E_J$ as already
seen in section IV.

\begin{figure}
\includegraphics[scale=0.3]{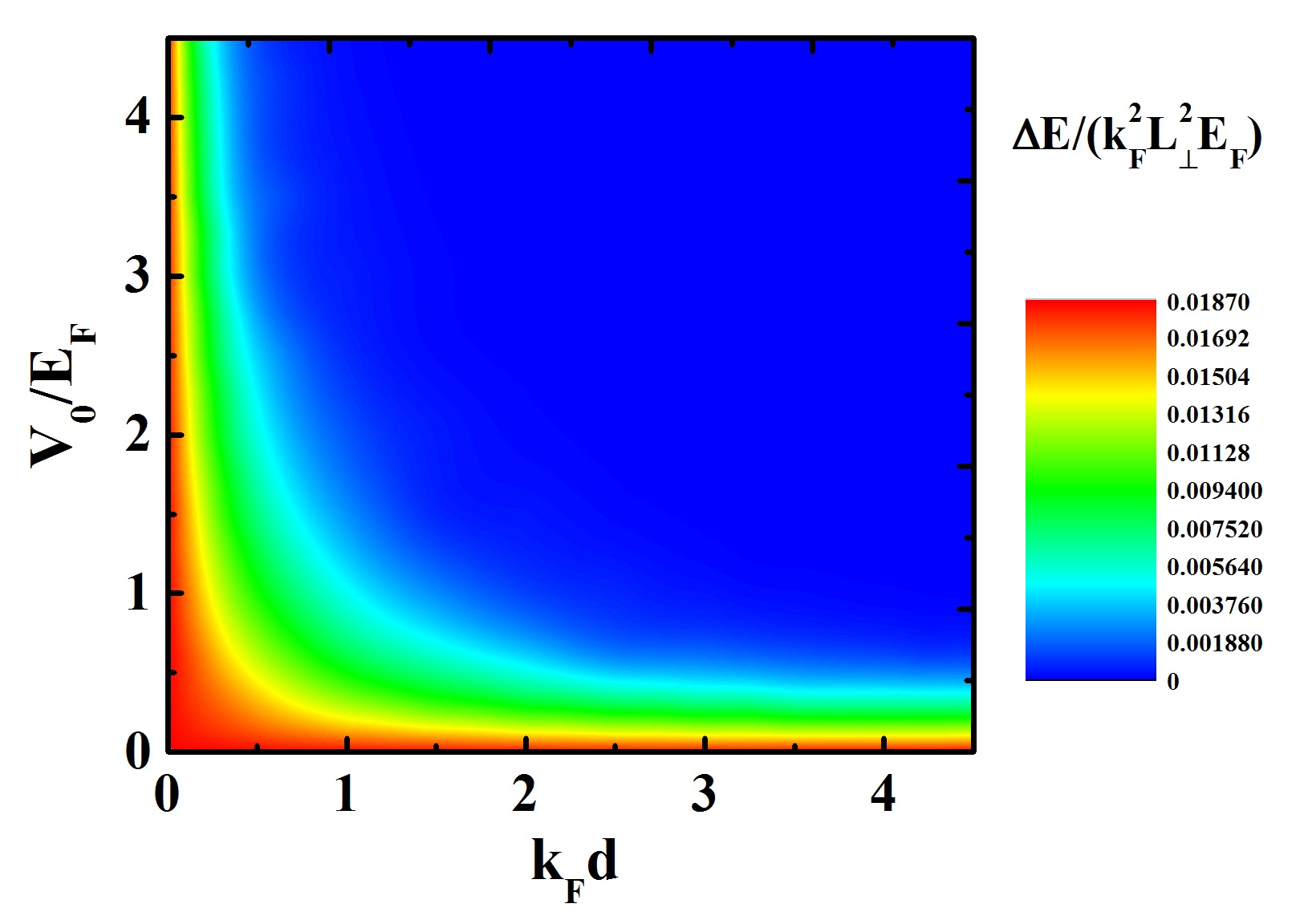}
\caption{Energy difference $\Delta E=E^--E^+$, divided by the area $L_\perp^2$,
as a function of the barrier height $V_0$ and width $d$. Here  $N=156$,
$L_{\perp}=13k_{F}^{-1}$, $L=30k_{F}^{-1}$, and $E_{\rm cut}=70E_{F}$.}
\label{figEjVd-d2v2}
\end{figure}
\begin{figure}
\includegraphics[scale=0.3]{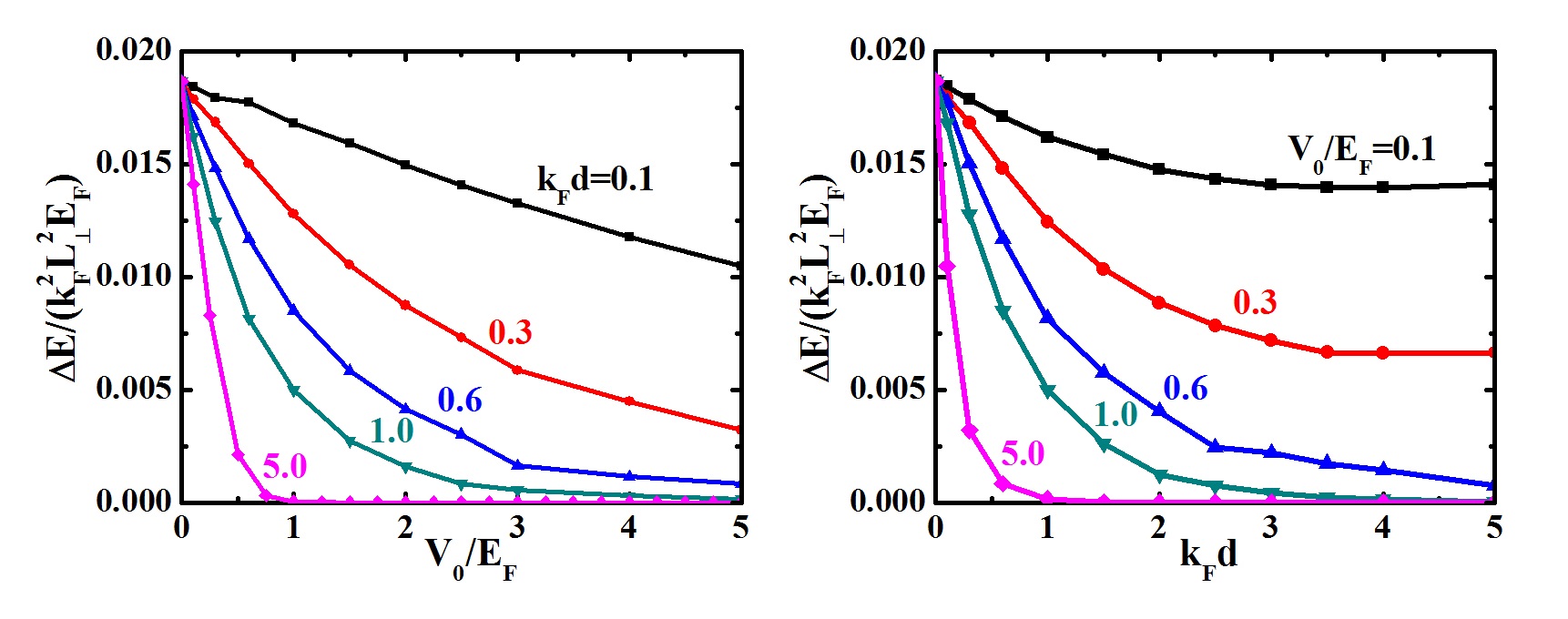}
\caption{Left: energy difference $\Delta E=E^--E^+$, divided by the area $L_\perp^2$, as
a function of the barrier height $V_0$ for different widths $d$. Right:
the same quantity as a function of the barrier width $d$ for different values
of height $V_0$. All the other parameters are the same as in the
previous figure. }
\label{figEjVd}
\end{figure}

Since the quantity $\Delta E$ is typically much smaller than the energies
$E^-$ and $E^+$, we must take care of  all possible sources of numerical
inaccuracy, in particular those introduced by the finite cutoff energy and the
finite box. We have first checked that, at unitarity, without barrier and in the
limit of large $L$,  the energy $E^+$ converges to the energy of a uniform infinite
gas: $(E^+/N)= (3/5)\mu= (3/5)(1+\beta)E_F= 0.354 E_F$, where $\beta=-0.41$
is the value of the Bertsch parameter in BdG theory \cite{rmp08}. In the same
situation, the quantity $\Delta E$ measures the cost in energy associated to
creation of the density depletion and the nodal structure of the order parameter
at the box center, corresponding to a dark soliton \cite{antezza}. The energy
per unit surface of a planar dark soliton at unitarity is $\epsilon_s=\Delta E/
(E_F k_F^2 L_\perp^2)$ and BdG theory gives  $\epsilon_s=(1+\beta)^{1/2}/(8\pi\sqrt{3})
\simeq 0.0176$ \cite{brand}. We have checked that both analytic values are
reproduced within an accuracy of about $2$\%, which is enough for our purposes.

Our results for $\Delta E$ as a function of $V_0$ and $d$ are shown in
Figs.~\ref{figEjVd-d2v2} and \ref{figEjVd}. All results in these figures
are obtained by using $N=156$, $L_{\perp}=13k_{F}^{-1}$, $L=30k_{F}^{-1}$,
and $E_{\rm cut}=70E_{F}$. As expected, $\Delta E$ approaches the same value
in the limit of vanishingly small barrier (i.e, for $V_0 \to 0$ at finite $d$).
This value is $\Delta E/(k_F^2L_\perp^2E_F)\simeq 0.0185$, which is slightly
larger than the energy of  dark soliton in an infinite system, due to the finite
box sixe. For $d$ of the order of $k_F^{-1}$, the quantity $\Delta E$
is rapidly decreasing when $V_0$ increases. The case of  large barriers
and small tunnelling (weak link) is where the physics of the Josephson
effect is expected to manifest.

Finally we note that for $V_0$ much smaller than $E_F$ the quantity $\Delta E$
tends to be a constant value when $d \to \infty$. A simple explanation
of this behavior is obtained by considering that, for a very wide
and low barrier, the effect of the barrier is that of lowering the
``bulk'' density in the central region of the box.  If $d$ is larger
that the soliton width, which is of the order of a few $k_F^{-1}$,
this effect can be accounted for by calculating the energy of a
dark soliton in uniform gas of reduced density.
Further increasing the width of the barrier has no effects on the soliton
energy and hence $\Delta E$ remains constant.


\begin{thebibliography}{99}
\bibitem{noteN}  Here the notation is the same as in Ref.\protect\cite{adhikari} except for the meaning of
$N$ which, in the present work, is the number of fermionic atoms, while in \protect\cite{adhikari} is
the number of bosonic dimers. In the context of NLSE the two numbers differ only by a factor two.
\end{thebibliography}
\end{document}